\documentstyle[12pt]{article}
\input epsf.tex

\renewcommand{\theequation}{\thesection.\arabic{equation}}
\newcommand{\be}{\begin{equation}}   \newcommand{\ee}{\end{equation}}
\newcommand{\bear}{\begin{eqnarray}}
\newcommand{\eear}{\end{eqnarray}}
\newcommand{\ba}{\begin{array}}      \newcommand{\ea}{\end{array}}
\newcommand{\lae}{\begin{array}{c}\,\sim\vspace{-21pt}\\< \end{array}}

\begin{document}

\pagestyle{empty}
\begin{titlepage}
\def\thepage {}        

\title{\Large \bf
Minimal Composite Higgs Model with Light Bosons}

\author{\bf \large  Bogdan A.~Dobrescu  \\
\\
{\small {\it Fermi National Accelerator Laboratory}}\\
{\small {\it P.O. Box 500, Batavia, Illinois 60510, USA \thanks{e-mail
  address: bdob@fnal.gov} }}\\ }

\date{ }

\maketitle

  \vspace*{-7.9cm}
\noindent
\makebox[10.7cm][l]{hep-ph/9908391} FERMILAB-PUB-99-234-T \\ [1mm]
\makebox[10.7cm][l]{August 17, 1999}
\\

 \vspace*{9.3cm}

\baselineskip=18pt

\begin{abstract}

   {\normalsize
We analyze a composite Higgs model
with the minimal content that allows a light Standard-Model-like 
Higgs boson, potentially just above the current LEP limit. 
The Higgs boson is a bound state made up of the top quark and a heavy 
vector-like quark.
The model predicts that only one other bound state may be lighter than
the electroweak scale, namely a CP-odd neutral scalar.
Several other composite scalars are expected to have masses in the TeV range.
If the Higgs decay into a pair of CP-odd scalars is kinematically open,
then this decay mode is dominant, with important implications for Higgs 
searches.
The lower bound on the CP-odd scalar mass is loose, 
in some cases as low as $\sim$ 100 MeV,
being set only by astrophysical constraints.}

\end{abstract}

\vfill
\end{titlepage}

\baselineskip=18pt
\pagestyle{plain}
\setcounter{page}{1}

\tableofcontents

\vspace*{4mm}

\section{Introduction}

The Standard Model is phenomenologically successful 
as an effective theory below some energy scale where 
new degrees of freedom (other than the yet-to-be-discovered Higgs boson)
should become relevant. Generic evidence for new 
physics is provided by the unphysical Landau poles for the quartic,
hypercharge and Yukawa couplings within the Standard Model, and by the 
existence of the gravitational interactions.
Barring an unlikely tuning of the parameters, the scale  $M_c$
of new physics that has an impact on the Higgs self-energy should
be in the TeV range or below. The nature of this new physics remains
unknown, and until experimental evidence for physics beyond the Standard 
Model will emerge, we should seek plausible explanations or alternatives
to the less compelling aspects of the Standard Model. 
One such aspect is that the Higgs doublet is an ad-hoc part of the 
Standard Model, which fits well the data but does not have an intrinsic
motivation. This remains true for supersymmetric or grand unified extensions
of the Standard Model. By contrast, the fermion content of the Standard Model
is better motivated, due to the anomaly cancellations and chiral symmetries.

It is therefore useful to
investigate the possibility that the Higgs doublet is not a fundamental 
degree of freedom but rather a bound state, that appears only in the 
effective theory below the scale $M_c$. 
Composite models in which the Higgs doublet is 
made up of some new fermions which belong to chiral representations of
the electroweak gauge group have been known for a long time 
\cite{Kaplan:1984sm}. 
Currently, the electroweak precision 
measurements constrain tightly the number of new chiral fermions, 
so that this type of models is disfavored unless there are 
non-perturbative effects or other phenomena that reduce the deviations
of the electroweak observables.

An economical way of satisfying the constraints from the electroweak data
is to bind a Higgs doublet out of the known fermions.
The top quark, having the mass close to the electroweak scale, is a prime
candidate for a Higgs constituent. 
However, if the Higgs doublet were a $\bar{t}_R t_L$ bound state, then 
a fairly reliable relation between the top quark mass, $m_t$, 
and the electroweak scale, $v \approx 246$ GeV, can be derived 
\cite{Bardeen:1990ds, Nambu, Cvetic:1997eb}. 
Given the measured value $m_t \approx 175$ GeV, models of this type 
could produce sufficiently large $W$ and $Z$ masses only if the 
compositeness scale is exponentially larger than the electroweak scale, 
and therefore they require fine-tuning.

Thus, it appears necessary that some new states play the 
role of Higgs constituents. A minimal choice is to introduce a vector-like 
quark, $\chi$, and non-perturbative four-quark interactions that involve 
the $\chi$ and $t$. Consequently,
the vacuum becomes populated with $\bar{\chi}_R t_L$ virtual pairs 
which make it opaque to the $W$ and $Z$, so that the electroweak symmetry is 
broken. Furthermore, the $t$ and $\chi$ mix, allowing $m_t \approx 175$ GeV
and a $\chi$ mass in the TeV range. 
This is the top condensation seesaw mechanism \cite{Dobrescu:1998nm}.
Below the scale of the four-quark operators, the 
effective theory contains a number of composite scalars, including a CP-even
neutral Higgs boson which is mainly a $\bar{\chi}_R t_L$ bound state
\cite{Chivukula:1998wd}. This theory has a decoupling 
limit in which at low energy it behaves as the Standard Model, 
and therefore is phenomenologically viable.

The four-quark interactions should be softened at high energy 
within a renormalizable or finite theory. Examples of this type 
involve new spontaneously broken gauge symmetries 
\cite{Dobrescu:1998nm, Chivukula:1998wd, family}
or extra dimensions accessible to the gluons 
\cite{Dobrescu:1998dg, Dobrescu:1999cs}.

In this paper we study in detail a minimal composite 
model which allows a light Standard-Model-like Higgs boson.
This model is based on the top condensation seesaw mechanism,
and the groundwork for its analysis is the effective potential formalism 
presented in Ref.~\cite{Chivukula:1998wd}.
Here we focus on the low-energy effective theory and its phenomenological
implications. 

In Section 2 we discuss the compositeness condition, and we 
identify a minimal set of ingredients necessary for the existence
of a light composite Higgs boson.

In Section 3 we write down the effective potential, and we discuss the
Higgs boson spectrum.
We establish that, besides the lightest neutral 
CP-even scalar, the composite Higgs sector may include only one 
physical state below a scale of order 1 TeV.
This is a CP-odd  scalar, which we will generically call the composite 
axion. 

The properties of the composite axion and lightest neutral 
CP-even Higgs boson are presented in Section 4. In Section 5 we 
turn to the Higgs couplings to the quarks and leptons.

In Section 6 we study the lower bounds on the  composite 
axion mass. In Section 7 we compare the Minimal Composite Higgs Model
with the Minimal Supersymmetric Standard Model, and we make some final 
remarks on phenomenology. In the Appendix we 
list the extremization conditions for the effective potential.

\section{Ingredients of a Minimal Composite Higgs Model}
\setcounter{equation}{0}

The Higgs sector of the Standard Model depends mainly on three parameters:
the Higgs doublet squared-mass, $M_H^2$, the quartic coupling, $\lambda$,
and the top Yukawa coupling, $y_t = m_t\sqrt{2}/v \approx 1$.
The relevant piece of the Lagrangian at the electroweak scale is given by 
\be
{\cal L}_{\rm SM}(v) = (D^\nu H^\dagger)(D_\nu H) - M_H^2(v) H^\dagger H
- \frac{\lambda(v)}{2} (H^\dagger H)^2
- y_t \left(\overline{\psi_L^3} t_R H + {\rm h.c.}\right) ~,
\ee
where the Lagrangian is defined at the electroweak scale, $\psi_L^3$
is the top-bottom left-handed doublet, and
we have chosen $H$ to have hypercharge +1 for convenience.
The other couplings are of little relevance for the 
renormalization group evolution of these three
parameters (one possible exception 
would be large neutrino Yukawa couplings, which in the presence 
of large Majorana masses yield acceptable neutrino masses;
we will not consider this possibility here).

If the Higgs doublet is a bound state with a compositeness scale $M_c$,
then at scales above $M_c$ the Higgs is no longer a physical degree of freedom.
Therefore, its kinetic term should vanish at $M_c$.
We will refer to this requirement as the compositeness condition
\cite{Bardeen:1990ds,Bando:1991rr}.
Note that this is equivalent with the statement that all the Higgs parameters
blow up at $M_c$ if the kinetic term normalization is fixed.

\subsection{The vector-like quark}

The top quark loop correction to the Higgs kinetic term is negative, 
diminishing the wave function renormalization,
suggestive of the possibility that the Higgs doublet is a $\bar{t}_R t_L$
bound state \cite{Bardeen:1990ds, Nambu, Cvetic:1997eb}. 
However, the top Yukawa coupling is perturbative, $y_t \approx 1$, 
and the kinetic term may vanish only if the $M_c$ scale is exponentially 
higher than the electroweak scale, such that the logarithm overcomes 
the loop factor.
Although a large hierarchy between  $M_c$ and the electroweak scale cannot
be ruled out, we will ignore this possibility due to a lack of explanation 
of the exponential fine-tuning required in that case.

Therefore, the compositeness condition requires new physics 
above the electroweak scale in order to speed up the running of the 
Higgs parameters. A simple choice is to include a vector-like quark, $\chi$,
which has the same transformation properties under the Standard
Model  gauge group as $t_R$,
and a mass $m_\chi > v$. This introduces a new Yukawa coupling:
\be
- \xi \left(\overline{\psi}_L^3 \chi_R H + {\rm h.c.}\right) ~.
\ee
If $\xi $ is sufficiently large, then the $\chi$ contribution to the 
Higgs self-energy may lead to the
cancellation of the Higgs kinetic term at a scale $M_c$ 
which is not hierarchically
bigger than $m_\chi$. However, in this case the renormalization group evolution
is nonperturbative (the cancellation of the Higgs kinetic term requires
the loop expansion parameter to be of order one), and a very precise 
computation is not within reach. Fortunately, the $\chi$ is a color triplet,
so that we can use an expansion in $1/N_c$, where $N_c$ is the number of 
colors. 
In this case the leading effects of $\chi$ on the Higgs parameters 
are the same as the perturbative one-loop contributions.
Although it is hard to estimate precisely how large are the corrections 
from the non-leading-$N_c$ terms, trading the 
physical problem of fine-tuning $M_c/m_\chi \gg 1$ for the
computational problem at $M_c \sim m_\chi$ seems justified. 
In practice, these two problems may be balanced by considering a
small hierarchy between $M_c$ and $m_\chi$, such that the 
fine-tuning is not excessive while the $\xi$ Yukawa coupling is not
much larger than one.

At a scale $\mu > m_\chi$, the Higgs sector takes the form
\bear
{\cal L}_{{\rm SM} + \chi}(\mu) & = & Z_H(\mu)(D^\nu H^\dagger)( D_\nu H) - 
M_H^2(\mu) H^\dagger H
- \frac{\lambda(\mu)}{2} (H^\dagger H)^2
\nonumber \\ [3mm]
&& - \left[ \overline{\psi}_L^3 \left(y_t t_R + \xi \chi_R \right) H
+  m_\chi \overline{\chi}_L\chi_R + {\rm h.c.}\right] ~,
\eear
while below $m_\chi$ the $\chi$ is integrated out and we recover the Standard 
Model.

\begin{figure}[t]
\begin{picture}(388,140)(-45,-20)
\thicklines
\put(100,50){\circle{40}}
\put(80,50){\line(-2, 0){11}}
\put(63,50){\line(-2, 0){11}}
\put(46,50){\line(-2, 0){11}}
\put(95,15){$\psi^3_L$}
\put(35,55){$H$}
\put(120,50){\line(2, 0){11}}
\put(137,50){\line(2, 0){11}}
\put(154,50){\line(2, 0){11}}
\put(93,77){$\chi_R$}
\put(155,55){$H$}
\thicklines
\put(300,50){\circle{40}}
%
\put(286,64){\line(-1,1){10}}
\put(272,78){\line(-1,1){10}}
\put(258,92){\line(-1,1){10}}
\put(293,77){$\chi_R$}
\put(293,17){$\chi_R$}
\put(262,45){$\psi^3_L$}
\put(325,45){$\psi^3_L$}
\put(286,36){\line(-1,-1){10}}
\put(272,22){\line(-1,-1){10}}
\put(258,8){\line(-1,-1){10}}
\put(314,36){\line(1,-1){10}}
\put(328,22){\line(1,-1){10}}
\put(342,8){\line(1,-1){10}}
\put(314,64){\line(1,1){10}}
\put(328,78){\line(1,1){10}}
\put(342,92){\line(1,1){10}}
\end{picture}
\caption[]{
\label{Figure1}
\small  Large-$N_c$ contributions to the Higgs doublet self-energy and 
quartic coupling.}
\end{figure}
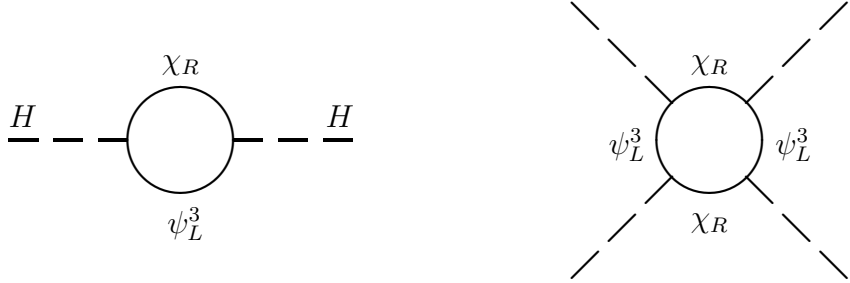
A straightforward computation of the one-loop Higgs self-energy 
and quartic coupling (see Fig.~1) gives
\bear
Z_H(\mu) & = & 1 -  
\frac{ N_c \xi^2}{16 \pi^2} 
\ln \left( \frac{\mu^2}{m_\chi^2} \right) ~,
\nonumber \\ [3mm]
\lambda(\mu) & = & \lambda(v) + 2\xi^2
\left[Z_H(\mu) -1 \right] ~,
\nonumber \\ [3mm]
M_H^2(\mu) & = & M_H^2(v) + 
\frac{N_c\xi^2}{8 \pi^2} \left(\mu^2 - v^2 \right)  ~,
\eear
where we neglected the top-quark contributions.
The compositeness condition, $Z_H(M_c) = 0$, yields 
\be
\xi^2 = \frac{8\pi^2}{N_c\ln\left(M_c/m_\chi\right)} ~.
\label{xi}
\ee
Since the ratio $M_c/m_\chi$ is unlikely to be exponentially large, 
it follows that $\xi \gg  1$, suggesting that the Higgs doublet
is mainly a $\overline{\chi}_R \psi^3_L$ bound state. But as stated before, 
keeping a reasonably small hierarchy  between $M_c$ and $m_\chi$ 
allows more control over the computation. For example, $M_c/m_\chi \sim
10 - 100$ gives $\xi \sim 3.4 - 2.4$.

If we impose $\lambda(\mu) > 0$ at all scales below $M_c$, so that 
the scalar potential is bounded from below, then
the quartic coupling at the electroweak scale,
\be
\lambda(v) = \lambda(M_c) + 2 \xi^2 ~,
\label{lambda}
\ee
is significantly larger than one, corresponding to a 
large Higgs boson mass, $v \sqrt{\lambda(v)}$.
After the non-leading contributions (from finite-$N_c$, top quark,
electroweak, and QCD effects)
 are taken into account, we expect the  Higgs boson mass to 
be close to the unitarity bound of 0.7 -- 0.9 TeV.

\subsection{Extending the Higgs sector}

So far we have shown that the compositeness condition, $Z_H(M_c) = 0$,  
suffices
to prove that the Higgs doublet cannot be a $\overline{t}_R \psi_L^3$ bound
state without exponential fine-tuning, while it can be a 
$\overline{\chi}_R \psi_L^3$ bound state provided the Higgs boson
is quite heavy. 

Next we would like to identify the circumstances which allow 
the composite Higgs boson to be light, close to the current experimental 
bounds. 
The large quartic coupling is a rather generic feature of 
a composite Higgs sector. However, only in the Standard Model
the Higgs boson mass is straightforwardly
determined by the  quartic coupling. For
extended Higgs sectors, the mixing between different CP-even scalars may drive
the lightest neutral Higgs boson significantly below the Standard Model 
unitarity bound. 
In order to allow a large scalar mixing, the constituents of the composite
Higgs sector should mix themselves. For the minimal fermion content, 
{\rm i.e.} three generations of quarks and leptons plus the vector-like quark
 $\chi$, the only fields that may have large mixings with the
$\chi_R$ and $\psi_L^3$ are the $t_R$ and $\chi_L$.

Therefore, we will consider a composite Higgs sector which involves 
four scalar fields: two
weak-doublets, $H_\chi \sim \overline{\chi}_R \psi_L^3$ and
$H_t \sim \overline{t}_R \psi_L^3$, and two weak-singlets,
$\phi_{\chi t} \sim \overline{t}_R \chi_L$ and 
$\phi_{\chi\chi} \sim \overline{\chi}_R \chi_L$.
Note that the case where one of the fermion fields is not 
a Higgs constituent can be recovered by taking the masses of the 
corresponding two scalars to infinity, but in that case
the Higgs boson is heavy \cite{Dobrescu:1999cs}.

For an extended Higgs sector, a natural formulation of 
the compositeness condition 
is that all scalar kinetic terms vanish at the same scale.
In the large-$N_c$ limit, the only contribution to a scalar kinetic term
comes from the fermions with large Yukawa couplings to that composite
scalar, namely from its constituents. 
Hence, the chiral symmetry of the constituents, $U(3)_L\times U(2)_R$,
is preserved by the Yukawa couplings of the composite scalars:
\be
\xi \left( \overline{\psi}_L^3 \, , \; \overline{\chi}_L \right) 
\Phi 
\left( \begin{array}{c}  t_R  \\ [3mm] \chi_R \end{array} \right)
+ {\rm h.c.} 
\label{yukc}
\ee
where the scalar $\Phi$ is a $3\times 2$ complex matrix,
\be
\Phi = 
\left( \begin{array}{cc} H_t  & - H_\chi \\ [3mm] 
 \phi_{\chi t} &  \phi_{\chi\chi}
 \end{array} \right) ~,
\label{yukawa}
\ee
with the phase of $H_\chi$ chosen negative for later convenience.
Note that the $SU(2)_W \times U(1)_Y$
electroweak symmetry is a gauged subgroup of this chiral symmetry.

Likewise, the leading-$N_c$ contributions to the running of the 
quartic couplings between the scales $v$ and $\mu$ is $U(3)_L\times U(2)_R$
symmetric:
\be
{\cal L}_{\rm quartic}(\mu) = {\cal L}_{\rm quartic}(v) 
- \frac{\lambda(v)}{2}
{\rm Tr} \left[\left( \Phi^\dagger \Phi \right)^2 \right] ~.
\ee
There are no other $U(3)_L\times U(2)_R$ symmetric terms in the scalar
potential. 

Since the $\chi$ quark is vector-like, and transforms under the Standard Model
gauge group as the $t_R$, we can write 
two gauge invariant mass terms:
\be
\mu_{\chi t} \overline{\chi}_L t_R 
+ \mu_{\chi \chi }\overline{\chi}_L \chi_R + {\rm h.c.} 
\ee
These break explicitly the chiral symmetry down to 
$SU(2)_W \times U(1)_Y\times U(1)_B$, where the last group refers to 
a global baryon number.
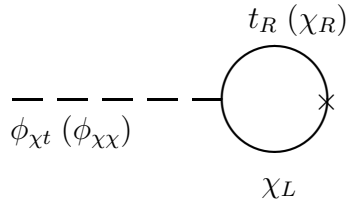
\begin{figure}[ht]
\centering
\begin{picture}(198,100)(-45,10)
\thicklines
\put(100,50){\circle{40}}
\put(115,46){$\times$}
\put(80,50){\line(-2, 0){11}}
\put(63,50){\line(-2, 0){11}}
\put(46,50){\line(-2, 0){11}}
\put(29,50){\line(-2, 0){11}}
\put(12,50){\line(-2, 0){11}}
\put(95,15){$\chi_L$}
\put(0,35){$\phi_{\chi t} \; (\phi_{\chi\chi})$}
\put(90,77){$t_R \; (\chi_R)$}
\end{picture}
\caption[]{
\label{Figure}
\small  Tadpole terms for the electroweak singlet scalars.}
\end{figure}
The effect of these explicit mass terms is to induce tadpole terms for 
the weak-singlet scalars in the effective potential 
(see Fig.~2):
\be
- \left(C_{\chi t} \phi_{\chi t} + C_{\chi\chi}\phi_{\chi\chi} 
+ {\rm h.c.}\right) ~.
\ee 
The tadpole coefficients may be estimated by cutting-off the loop integral 
at $M_c$. For $\mu_{\chi t, \, \chi\chi} \ll M_c$,
\be
C_{\chi t, \, \chi\chi} \approx \frac{N_c \xi}{8\pi^2} 
\mu_{\chi t, \, \chi\chi} M_c^2 ~.
\label{param}
\ee
Another effect of these explicit mass terms is to induce
 trilinear scalar terms proportional with 
$\mu_{\chi t, \, \chi\chi}$, due to the large-$N_c$ running 
between the scales $v$ and $M_c$. 

A generic high energy theory at the scale $M_c$ gives rise to the 
most general mass terms for the composite scalars, which 
also break explicitly the $U(3)_L\times U(2)_R$ chiral symmetry down to 
$SU(2)_W \times U(1)_Y\times U(1)_B$.
Putting together all these terms, the scalar potential for the 
two-doublet-two-singlet composite Higgs sector has all possible 
gauge invariant terms and is hard to analyze. In order to progress 
we need to make some assumptions about the high energy theory that is 
responsible for binding together the $\chi$, $t_R$ and $\psi^3_L$
within the composite scalars.

First, we can invoke a small hierarchy between 
the compositeness scale and the masses of the composite scalars, as mentioned 
in section 2.1. As a result, the trilinear, quartic and higher-dimensional 
Higgs couplings at the $M_c$ scale are small, suppressed by powers
of the $M_c/m_\chi$ ratio.
Second, we will see in section 3.3 that the sector of
the high-energy theory responsible for binding the composite Higgs sector
is likely to preserve a global 
$U(1)_{t} \times U(1)_{\chi} \times U(1)_B$ subgroup
of the chiral symmetry. This symmetry precludes the presence of
mass terms that mix the doublets or the singlets 
in the effective potential.
 The quark charges under this symmetry are 
 determined only up to a unitary transformation. A simple basis is  
that where only $t_R$
and $\chi_R$ are charged under $U(1)_{t}$ and $U(1)_{\chi}$, respectively. 
One linear combination of these two $U(1)$'s has an axial QCD anomaly,
but this effect may be neglected as we will argue in Section 6.

With these assumptions, one can easily integrate out the composite 
scalars at scales above $M_c$, where they can be treated as 
non-propagating (spurion) fields.
This bottom-up approach results in the following four-quark operators
at the scale $M_c$:
\begin{equation}
{\cal L}_{\rm eff} = \frac{g_{\psi\chi}^2}{M_c^2} 
\left(\overline{\psi}_L^3 \chi_R \right)
\left(\overline{\chi}_R \psi_L^3  \right)
+ \frac{g_{\psi t}^2}{M_c^2} \left(\overline{\psi}_L^3 t_R \right)
\left(\overline{t}_R \psi_L^3 \right) 
+ \frac{g_{\chi t}^2}{M_c^2} \left(\overline{\chi}_L t_R \right)
\left(\overline{t}_R \chi_L \right) 
+ \frac{g_{\chi\chi}^2}{M_c^2} \left(\overline{\chi}_L \chi_R \right)
\left(\overline{\chi}_R \chi_L \right) 
\label{fourq}
\end{equation}
Altogether,
there are seven parameters: the four coefficients of
the above operators, the two masses ($\mu_{\chi\chi}$ and $\mu_{\chi t}$),
and the overall scale $M_c$.
The effective potential below the scale $M_c$ is sufficiently simple to 
be analyzed analytically. Before doing so in Section 3, we will argue in 
the remainder of this Section that the assumptions made here are realistic.

\subsection{Candidates for physics above the compositeness scale}

The basic assumption we are making for an extended composite 
Higgs sector is that all scalar kinetic terms vanish at the 
same scale, referred to as the compositeness scale $M_c$.
Therefore, the composite scalars are no longer physical degrees of freedom
and they should be integrated out above $M_c$. This gives rise to 
higher-dimensional operators which at high-energy should be 
replaced by a renormalizable or finite theory.

A conspicuous direction for seeking such a high-energy theory 
is to consider some new gauge dynamics which binds the $t$ and $\chi$
within the composite scalars. Such dynamics cannot be confining because
the top has already been observed by the CDF and D0 collaborations.
On the other hand, the new gauge interactions have to be rather strongly 
coupled at the compositeness scale in order to deeply bind the Higgs doublets 
and  trigger the electroweak phase transition.
Therefore, unless the new physics is very unconventional right above
the compositeness scale, the new gauge interactions must be asymptotically 
free. These requirements single out spontaneously broken non-Abelian 
gauge theories.

The choice of a gauge group is further restricted if no new chiral generations
of fermions are introduced. The representations of the new non-Abelian 
gauge group may coincide with those of $SU(3)_C$, as in topcolor 
\cite{topcolor}, or may correspond to some flavor or family symmetry
\cite{family}.

The compositeness scale is approximately given by the masses of the
heavy gauge bosons. Below $M_c$, the gauge bosons are integrated out 
resulting in higher-dimensional operators, including those listed in 
eq.~(\ref{fourq}). The four-quark operators with left-left or 
right-right current-current 
structure do not contribute to the effective potential in the large-$N_c$ 
limit (though they do contribute to observables, most importantly to
the $\rho$ parameter \cite{Chivukula:1995dc, Chivukula:1998uf}, 
but these contributions are sufficiently small 
for $M_c$ above a few TeV \cite{Burdman:1999us}). 
All other four-quark operators which are invariant under the 
Standard Model gauge group and involve only the $\psi_L$, $t_R$
and $\chi$ fields violate the global $U(1)_{t} \times U(1)_{\chi}$ symmetry, 
and are not expected to be induced by heavy gauge boson exchanges. 

Operators of dimension-8 or higher are also induced by the gauge dynamics.
However, their effects are negligible at scales significantly below $M_c$.
Therefore it is convenient to ignore them by arranging 
a small hierarchy between $M_c$ and the composite scalar masses.
Such a hierarchy arises if there is a second order phase transition
in which a continuous variation of
the gauge coupling induces a continuous variation of the 
scalar masses. 
There are various arguments, based in general on the large-$N_c$ limit,
indicating that a spontaneously broken gauge 
group leads indeed to a second order phase transition \cite{Chivukula:1998uf,
 Collins:1999cf}. In practice, since
we do not require an exponential hierarchy, it is sufficient to
have a weakly first order phase transition.

The new gauge dynamics should be flavor dependent, so that only 
the top and $\chi$ acquire large masses. This can be realized in various
ways. The strongly coupled gauge interaction may act only on the 
third generation quarks and on the $\chi$, while the splitting between the
$\chi$, $t$, $b$ masses may be given by some perturbative interactions.
Examples of this type have been given in  
\cite{Dobrescu:1998nm, Chivukula:1998wd}. 
Alternatively, the strongly coupled gauge interaction may be flavor universal,
with the flavor breaking provided by an extended vector-like quark sector
\cite{family}. 

Above the $M_c$ scale there must be some additional 
physics that leads to the spontaneously breaking of the non-Abelian 
gauge symmetry responsible for Higgs compositeness. This may involve 
new gauge dynamics, or fundamental scalars and supersymmetry.
Yet another alternative may be provided by quantum gravitational effects 
if gravity is modified at short distance \cite{largedim,
Randall:1999ee, Lykken:1999nb} such that it becomes
strong at a scale in the multi-TeV range, not far above $M_c$.

Instead of a new gauge symmetry, the binding of the Higgs sector 
may  be produced by the Standard Model gauge bosons propagating 
in extra dimensions \cite{Dobrescu:1998dg,Dobrescu:1999cs} of radius 
$1/M_c$.
Basically, the exchange of Kaluza-Klein modes of the gluons
induces  four-quark operators of the type (\ref{fourq}).
Although the Kaluza-Klein modes are weakly coupled at a TeV scales, the
combined effect of all modes is nonperturbative.
This is also consistent with gauge coupling unification at a scale 
below 100 TeV \cite{Cheng:1999fu}, albeit the theoretical uncertainties are 
somewhat larger than in the Minimal Supersymmetric Standard Model.

Moreover, the extra dimensions  allow new explanations for flavor symmetry 
breaking, such as flavor-dependent positions of the fermions in the extra 
dimensions \cite{Arkani-Hamed:1999dc}, messengers of flavor-breaking
propagating in the bulk \cite{Berezhiani:1998wt}, or exponential suppressions
due to renormalization effects \cite{extra}.
The extra dimensions could also provide a natural reason for the 
existence of the vector-like quark, because
the $\chi$ has the same gauge quantum numbers as the 
Kaluza-Klein modes of the $t_R$. This idea however requires further study.

The gauge theories in extra dimensions are  non-renormalizable, 
so that new physics should
soften the interactions at a scale close to the compactification scale. 
This physics may be based on an underlying theory that includes quantum 
gravity, such as string or M theory.  Alternatively, a 
 physical cut-off to the interactions of the Kaluza-Klein modes
could be set by the brane recoil \cite{Bando:1999di}, potentially allowing the 
fundamental (string) scale to be substantially 
higher than the compactification scale.

\section{The Two-Doublet-Two-Singlet Higgs Sector}
\setcounter{equation}{0}

In this section we study 
the composite Higgs sector which includes two weak-doublets, $H_t$ and 
$H_\chi$, and two weak-singlets, $\phi_{\chi t}$ and $\phi_{\chi\chi}$.
The effective potential is determined based on
the following four assumptions discussed in section 2.2: \\
1. The compositeness condition: the kinetic terms of all composite scalars 
vanish at the same scale $M_c$. \\
2. There is a separation between the compositeness scale and
the scalar masses. \\
3. The interactions which bind the composite 
scalars preserve the $U(1)_t \times U(1)_\chi$ chiral symmetry of the 
$t_R$ and $\chi_R$ quarks. \\
4. The large-$N_c$ limit is a reasonable approximation for computing
the effects of the strong dynamics responsible for compositeness.

The effective potential below the compositeness scale is
given by
\bear
V & = & \frac{\lambda}{2} \left[
\left(H_t^\dagger H_t + \phi_{\chi t}^\dagger \phi_{\chi t}\right)^2
+ \left(H_\chi^\dagger H_\chi + 
\phi_{\chi\chi}^\dagger \phi_{\chi\chi}\right)^2 + 
2 \left| H_t^\dagger H_\chi -
\phi_{\chi t}^\dagger \phi_{\chi\chi}\right|^2 \right]
\nonumber \\ [2mm]
& & + M_{tt}^2 H_t^\dagger H_t + M_{t\chi}^2 H_\chi^\dagger H_\chi 
+ M_{\chi t}^2 \phi_{\chi t}^\dagger \phi_{\chi t} + 
M_{\chi\chi}^2 \phi_{\chi\chi}^\dagger \phi_{\chi\chi}
\nonumber \\ [2mm] & & 
+ \left(C_{\chi t} \phi_{\chi t} + C_{\chi\chi}\phi_{\chi\chi} 
+ {\rm h.c.}\right)
\label{efpot}
\eear
The seven parameters listed at the end of section 2.2 have been replaced 
by four real squared-mass parameters, two tadpole coefficients 
(chosen positive), and the quartic coupling.
With the exception of $\lambda$ which can be computed in the large-$N_c$
limit and depends only logarithmically on $M_c$, 
the other parameters are essentially free, and remain to be determined 
within the underlying theory above the compositeness scale.

The effective potential is $SU(2)_W \times U(1)_Y$ and CP invariant, and has 
a $U(1)_t \times U(1)_\chi$ global symmetry softly broken by the tadpole 
terms.
The tadpole terms also force the $\phi_{\chi t}$ and $\phi_{\chi\chi}$ to
have non-zero VEVs.
For $M^2_{t\chi} < 0$, there is a range of parameters
where the $H_\chi$ doublet has a non-zero VEV, breaking the 
electroweak symmetry. In that case, the 
third term of $V$ provides a tadpole for $H_t$, which acquires a VEV too. 
Finally, the condition $M^2_{tt} > 0$ is sufficient to keep the 
VEVs of the two doublets aligned, leaving the photon massless:
\bear
H_t & = &\left( \begin{array}{c} \frac{1}{\sqrt{2}} \left[ v \cos\beta
+ h^0_{tt} + i \left(A^0 \sin\beta - G^0 \cos\beta \right)\right]
\\ [4mm] H^-\sin\beta - G^-\cos\beta \end{array} \right) \; ~,
\nonumber \\ [5mm]
H_\chi & = & \left( \begin{array}{c} \frac{1}{\sqrt{2}} \left[ v \sin\beta
+ h^0_{t\chi} - i \left(A^0 \cos\beta + G^0 \sin\beta \right)\right]
\\ [4mm] - \left(H^-\cos\beta + G^-\sin\beta\right) \end{array} \right) \; ~,
\nonumber \\ [5mm]
\phi_{\chi t} & = & \frac{1}{\sqrt{2}} \left( 
- \frac{v \sin\beta}{\epsilon\tan\gamma}
+ h^0_{\chi t} + i A^0_{\chi t} \right) \; ~,
\nonumber \\ [4mm] 
\phi_{\chi\chi} & = & \frac{1}{\sqrt{2}} \left( - \frac{v}{\epsilon}\sin\beta 
+ h^0_{\chi\chi} + i A^0_{\chi\chi} \right)  \; ~,
\label{vacuum}
\eear
where we have written the VEVs in terms of the electroweak scale, fixed at
$v\approx 246 \; {\rm GeV}$, and three other parameters:
$\beta, \gamma \in (0, \;  \pi/2)$ and $\epsilon > 0$. 
These VEVs are related to the parameters in the effective potential by the 
extremization conditions listed in the Appendix. 
Note that the phases of the VEVs for $\phi_{\chi t}$ and $\phi_{\chi\chi}$
are fixed by the tadpole terms, the relative phase of the VEVs for 
$H_t $ and $H_\chi$ is fixed by the third term in Eq.~(\ref{efpot}),
and the phase of  $H_\chi$ has been chosen in the Yukawa couplings
(\ref{yukc}).

$G^\pm$ and $G^0$ are the Nambu-Goldstone bosons that become the longitudinal 
$W$ and $Z$. Altogether there are nine massive degrees of freedom, 
which are characterized by their electric charge and CP-parity: 
two charged states $H^\pm$, three CP-odd neutral scalars
$ A^0 , \; A^0_{\chi t}$, $A^0_{\chi\chi}$, and four
CP-even neutral scalars, $h^0_{tt}, \; h^0_{t\chi}$,
$h^0_{\chi t}$ and $h^0_{\chi\chi}$.
Before dissecting their spectrum, let us discuss the constraints on the 
parameter space.

The $H_\chi$ doublet contributes more to the electroweak symmetry breaking
than $H_t$ (this is the motivation for introducing the 
vector-like quark), so that $\tan \beta > 1$.
Due to the Yukawa couplings of the scalars to their constituents
[see Eq.~(\ref{yukc})], the $t$ and $\chi$ mix, with a mass matrix
\begin{equation}
\frac{\xi v \sin\beta}{\epsilon \sqrt{2}}
\left(\overline{t_L} ,\,\overline{\chi_L}\right)\,
\pmatrix{- \epsilon \cot\beta  & \epsilon \cr 
\cot\gamma & 1 \cr}
\pmatrix{t_R \cr \chi_R \cr} ~,
\label{f1}
\end{equation}
where the Yukawa coupling $\xi$ is given by Eq.~(\ref{xi}) 
in the large-$N_c$ limit. 
We are interested in the case where the $\chi$ is heavier than  the 
top, so that the corrections to the electroweak observables are small.
This implies $\epsilon < 1$. In what follows we will often consider the 
limit in which $\chi$ decouples, {\rm i.e.} $\epsilon \ll 1$.
The physical top quark is the light mass eigenstate of the above matrix:
\be
m_t \approx \frac{\xi v}{\sqrt{2}} \sin (\beta + \gamma) 
\left[ 1 + {\cal O}(\epsilon^2)\right] ~.
\label{top-mass}
\ee
Therefore, $\sin (\beta + \gamma) \approx 1/\xi$. 
For $\xi^2 \gg 1$
one has $\tan\beta, \tan\gamma \gg 1$. More generally, we allow
$\beta, \gamma \in (\pi/4 \, , \;  \pi/2)$, which also satisfies the 
$\cos (\beta + \gamma) < 0$ restriction imposed by 
the extremization condition written in the Appendix. 
Finally, the quartic coupling at the electroweak scale 
is related to the Yukawa coupling by 
$\lambda \approx 2 \xi^2$ in the large-$N_c$ limit, because the 
quartic coupling at the compositeness scale is assumed to be negligible
[see Eq.~(\ref{lambda})].

Let us proceed with the computation of the scalar spectrum.
The charged Higgs boson, $H^\pm$, has a mass
\be
M^2_{H^\pm} = \frac{\lambda}{2} v^2 
\left(\frac{\tan\beta}{\epsilon^2\tan\gamma} 
- 1\right) ~.
\ee
This sets the scale for the heavy composite scalars.
In addition, this is roughly the scale for the vector-like quark,
whose mass is given by
\be
m_\chi = M_{H^\pm} \sqrt{ \frac{\sin 2\beta}{2\, \sin 2\gamma}}
\left[1 + {\cal O}(\epsilon^2) \right] ~.
\ee

\subsection{CP-odd neutral scalars}

From the effective potential one can find the squared-mass matrix
for the three CP-odd neutral scalars,
$ A^0 , \; A^0_{\chi t}$, and
$A^0_{\chi\chi}$:
\be
\left(M^2_{H^\pm} + \frac{\lambda v^2}{2}\right)
\left( 1 + \epsilon^2 \frac{\cos^2\!\beta}{\cos^2\!\gamma} \right)
U_0 {\rm diag}\left(1,0,0\right) U_0^\dagger + 
\frac{\sqrt{2}\epsilon}{v\sin\beta}
 {\rm diag}\left(0, \, C_{\chi t} \tan\gamma, \,  C_{\chi\chi} \right) ~.
\ee
Up to an $SU(2)$ transformation, the matrix 
\be
U_0 = \left( \begin{array}{ccc} \cos\theta_1 \cos\theta_2 & 
- \sin\theta_1  &  \cos\theta_1 \sin\theta_2  \\ [3mm] 
\sin\theta_1 \cos\theta_2  & \cos\theta_1  & \sin\theta_1 \sin\theta_2  
\\ [3mm] 
- \sin\theta_2   &  0  &  \cos\theta_2
\end{array} \right) ~,
\ee
would define the mixing angles if the $C_{\chi t}$ and $C_{\chi\chi}$ 
were zero (we will see below that this would not be a viable case). 
The angles $\theta_1, \theta_2 \in (0, \pi/2)$ are given by
\bear
\tan\theta_1 & = & \epsilon \cos\beta \, \tan\gamma ~, 
\nonumber \\ [1mm] 
\tan\theta_2 & = & \epsilon \cos\beta \, \cos\theta_1 ~.
\eear

The mass matrix of the CP-odd states is diagonal in the
$\epsilon \rightarrow 0$ limit. Therefore, we can diagonalize it
by expanding in $\epsilon$, and we obtain the
following squared-masses of the CP-odd neutral scalars:
\bear 
M_{A^0_1}^2 & = & M^2_{H^\pm}
\left\{ 1 + \epsilon^2 \left[ 
\cos\beta \left( c_A \tan\gamma + c_A^\prime \right) 
+\frac{\tan\gamma} {\tan\beta}\right]
+ {\cal O}(\epsilon^4)\right\} ~,
\nonumber \\ [3mm] 
M_{A^0_t}^2 & = & C_{\chi t}\frac{\sqrt{2} \epsilon \tan\gamma}{v \sin\beta}
 \left[ 1 - \epsilon^2 c_A \tan\gamma \, \cos\beta
+ {\cal O}(\epsilon^4)\right] ~,
\nonumber \\ [3mm] 
M_{A^0_\chi}^2 & = & C_{\chi\chi}\frac{\sqrt{2} \epsilon}{v \sin\beta}
 \left[ 1 - \epsilon^2  c_A^\prime \cos\beta
 + {\cal O}(\epsilon^4)\right] ~.
\label{oddmass}
\eear
The two dimensionless coefficients, 
\bear
c_A & \equiv & \frac{\cos\beta \, \tan\gamma}{1 - 
\left( C_{\chi t}/C_0 \right) \tan\gamma } ~,
\nonumber \\ [3mm] 
c_A^\prime & \equiv  & \frac{\cos\beta}{1 - 
\left(C_{\chi\chi}/C_0 \right) } ~,
\eear
where
\be
C_0 \equiv \frac{\lambda v^3 }{2\sqrt{2} \epsilon^3} 
\frac{\tan\beta}{\tan\gamma}
\sin\beta ~, 
\ee
are defined such that the mass eigenstates of the CP-odd neutral scalars
take a simple form:
\bear 
A^0_1 & = & A^0 + \epsilon \left( c_A A^0_{\chi t}
- c_A^\prime A^0_{\chi\chi} \right)
+ {\cal O}(\epsilon^2)
\nonumber \\ [3mm] 
A^0_t & = & A^0_{\chi t} - \epsilon c_A A^0
+ {\cal O}(\epsilon^2)
\nonumber \\ [3mm] 
A^0_\chi & = & A^0_{\chi\chi} + \epsilon c_A^\prime A^0
+ {\cal O}(\epsilon^2)
\eear

The $A^0_1$ state is included predominantly in the Higgs doublets.
In the small-$\epsilon$ limit, it belongs to a linear combination of
Higgs doublets, namely $(-H_t \sin\beta + H_\chi \cos\beta)$, 
whose VEV vanishes.
The other states of this linear combination 
are the charged Higgs and a CP-even neutral scalar. 
The degeneracy of these states is lifted only by electroweak symmetry 
breaking effects. As a result, the mass splittings among these states are 
proportional with $v^2/M^2_{H^\pm} \sim \epsilon^2$, which explains the
first equation in (\ref{oddmass}).

The $A^0_t$ and $A^0_\chi$ are predominantly the imaginary parts of the 
weak-singlet fields. They are the Nambu-Goldstone bosons associated
with the $U(1)_t \times U(1)_\chi$ global symmetry 
of the effective potential, which is spontaneously 
broken at a scale of order $v/\epsilon$.
The tadpole terms from the 
effective potential break the $U(1)_t \times U(1)_\chi$ symmetry 
explicitly, so that the $A^0_t$ and $A^0_\chi$ acquire 
squared-masses proportional to $C_{\chi t}$ and $C_{\chi\chi}$, respectively.
This explains the second and third equations in (\ref{oddmass}).

\subsection{CP-even neutral scalars}

The composite Higgs sector includes four 
CP-even neutral scalars. The two states belonging to the Higgs doublets,  
$h^0_{tt}, h^0_{t\chi}$, do not mix in the $\epsilon \rightarrow 0$ limit
with the two states from the Higgs singlets, $h^0_{\chi\chi}$, 
$h^0_{\chi\chi}$. This allows us to identify immediately the mass 
eigenstates to leading order in $\epsilon$.
As a result we learn that 
it is convenient to write the squared-mass matrix
for CP-even neutral scalars  (without expanding in $\epsilon$) 
in the basis $(-\sin\beta\; h^0_{tt} + \cos\beta\; h^0_{t\chi})$,
$\; (\cos\beta\; h^0_{tt} + \sin\beta\; h^0_{t\chi})$,
$\; h^0_{\chi t}$ and $h^0_{\chi\chi}$:
\bear
{\cal M}^2_h & = & \frac{\lambda v^2}{2 \epsilon^2} 
\left( \begin{array}{cc}  
{\rm diag}\left(\frac{{\textstyle \tan\beta}}{{\textstyle \tan\gamma}}, \,  
2 \epsilon^2 \right) &  \epsilon {\cal B}
\\ [3mm] 
\epsilon {\cal B}^\top & 
\, \Gamma \, 
{\rm diag}\left(\frac{{\textstyle \sin 2\beta}}{{\textstyle \sin 2 \gamma}}
\, \epsilon^2 , \,  2 \frac{{\textstyle \sin^2\!\beta}}
{{\textstyle \sin^2\!\gamma}} \right) \, \Gamma^\top 
\end{array} \right)
\nonumber \\ [4mm]
&& + \frac{\sqrt{2}\epsilon}{v\sin\beta}
 {\rm diag}\left(0, \, 0, \, C_{\chi t} \tan\gamma, \,  C_{\chi\chi} \right)
\eear
where the $2\times 2$ matrix ${\cal B}$ depends only on $\beta$ and $\gamma$,
\be
 {\cal B} = \frac{\sin\beta}{\sin\gamma}
\left( \begin{array}{cc}  \sin (2\beta + \gamma) \, & \, 
\cos (2\beta + \gamma) \\ [3mm] 
-2 \cos\beta \, \cos (\beta  + \gamma) \, & \, 
2 \sin\beta \, \cos (\beta  + \gamma) 
\end{array} \right) ~,
\ee
and $\Gamma$ is a unitary matrix, 
\be
\Gamma = 
\left( \begin{array}{cc} \sin\gamma & \cos\gamma \\ [3mm] 
- \cos\gamma &  \sin\gamma 
 \end{array} \right) ~.
\label{Gamma}
\ee

From the mass matrix it can be seen that one of the mass eigenstates, $H^0_1$, 
 is predominantly the $(-\sin\beta\; h^0_{tt} + \cos\beta\; h^0_{t\chi})$
state.
Up to mixings of order $\epsilon$, $H^0_1$ forms together with $A^0_1$ and 
$H^\pm$ a weak-doublet with a zero VEV.
As discussed in the case of $A^0_1$, the mass splitting among these
states are given by electroweak 
symmetry breaking effects which show up in the heavy scalar spectrum only at 
order $\epsilon^2$:
\be
M^2_{H^0_1} = M^2_{H^\pm} \left[ 1 + {\cal O}(\epsilon^2) \right] ~.
\ee
This can also be checked directly from
the expression for ${\cal M}^2_h$. 

Two other scalars are linear combinations of the real
parts of the weak-singlets and order $\epsilon$ admixtures of the neutral
components of the weak-doublets. 
The third and fourth lines and rows of ${\cal M}^2_h$
give their masses to leading order in $\epsilon^2$:
\be
M^2_{H^0_{2,3}} =  M^2_{H^\pm}  \frac{\sin 2\beta} {\sin 2\gamma}
\left[1 + x_t + x_\chi \pm \sqrt{ \left( 1 + x_t - x_\chi \right)^2
- 4 \left(x_t - x_\chi \right)\sin^2\!\gamma } \,\right]
\left[ 1 + {\cal O}(\epsilon^2) \right] ~,
\ee
where we used the notation
\be
x_{t,\,\chi}  \equiv \frac{\sin 2\gamma}{2\sin 2\beta} \,
\frac{M^2_{A^0_{t,\,\chi}}}{M^2_{H^\pm}} > 0 ~.
\ee
It is straightforward to check that $M^2_{H^0_{2,3}}$ are positive 
everywhere within the parameter space. 

The only remaining mass eigenstate, which we label $h^0$, is 
predominantly $\; (\cos\beta\; h^0_{tt} + \sin\beta\; h^0_{t\chi})$.
Its mass cancels at leading order in $\epsilon^2$. 
This is due to the fact 
that $h^0$ sits mainly in the only combination of weak-doublets 
which breaks the electroweak symmetry. Hence, for fixed $v$
the $h^0$ mass does not depend on $\epsilon$, whereas the other 
scalar masses  scale as $1/\epsilon$.
To compute $M_{h^0}$ we have to go to the next-to-leading order in the 
$\epsilon$ expansion. Fortunately, we do not need to diagonalize
the $4\times 4$ ${\cal M}^2_h$  matrix. It is sufficient to 
compute the determinant of the mass matrix, and then to use
\be
{\rm Det} {\cal M}^2_h = M^2_{h^0} M^2_{H^0_1}M^2_{H^0_2}M^2_{H^0_3} ~.
\ee
The result is fairly simple:
\be
M^2_{h^0} =
\lambda v^2 \left[ 1 - \cos^2\!(\beta + \gamma) \frac{
x_t \sin^2\!\beta + x_\chi \cos^2\!\beta + \sin^2(\beta + \gamma) }
{x_t x_\chi + x_t\sin^2\!\gamma + x_\chi \cos^2\!\gamma} 
+ {\cal O}(\epsilon^2) \right]
\label{mhiggs}
\ee
The mass eigenstate corresponding to this eigenvalue 
may also be derived by expanding in powers of $\epsilon$:
\be
h^0 = \sin\beta\; h^0_{t\chi} + \cos\beta\; h^0_{tt} 
+ \epsilon \left( c_h h^0_{\chi t} + c_h^\prime h^0_{\chi\chi} \right)
+ {\cal O}(\epsilon^2) ~,
\ee
where the two dimensionless coefficients are defined by
\be
\left( \!\begin{array}{c} c_h \\ [3mm] c_h^\prime \end{array} \!\right)
= \frac{\sin\gamma \; \cos (\beta + \gamma)}
{ \sin\beta \left(
x_t x_\chi + x_t \sin^2\!\gamma + x_\chi \cos^2\!\gamma\right)}
\left[
\left( \!\begin{array}{c} x_\chi \cos\beta  \\ [3mm] 
- x_t \sin\beta \end{array} \!\right)
+ \sin (\beta + \gamma) \left( \!\begin{array}{c} \sin\gamma \\ [3mm] 
-\cos\gamma \end{array} \right) \!\right] ~.
\ee

We are now in a good position for discussing the vacuum stability.
The extremum conditions written down in the Appendix are 
automatically satisfied because they have been used
to replace the four mass-squared parameters from the effective potential
with four new parameters.
Therefore, the vacuum defined by eq.~(\ref{vacuum}) is a local minimum 
if and only
if all four eigenvalues of ${\cal M}^2_h$ are positive. We have seen that
three of them, $M^2_{H^0_1}, \, M^2_{H^0_2}$, and $M^2_{H^0_3}$ are 
always positive. The only remaining condition, $M^2_{h^0} > 0$, is
restrictive. For example, if both $M_{A^0_{t}}$ and $M_{A^0_{\chi}}$ 
were of order $v$ or lighter, then $x_t$ and $x_\chi$ were of order 
$\epsilon^2$, and $M^2_{h^0}$ would be negative.
Thus, at most one of $M_{A^0_{t}}$ and $M_{A^0_{\chi}}$ may be as light
as the electroweak scale, the other one having a mass of order $v/\epsilon$ 
or larger.

Imposing $M^2_{h^0} > 0$ ensures that the vacuum that we study is a local
minimum of the potential, but not necessarily a global minimum.
An inspection of the extremum conditions shows that there is only one 
other candidate for a global minimum, namely that obtained by taking 
$v \rightarrow 0$ and $v/\epsilon > 0$. 
This is easy to understand, because the 
tadpole terms always give rise to VEVs for $\phi_{\chi t}$ 
and $\phi_{\chi\chi}$. It is clear that the $v > 0$ minimum which we 
study here is deeper than the $v = 0$ extremum for sufficiently large 
and negative values of $M^2_{t \chi}$. It seems hard to compute analytically 
the critical value for $M^2_{t \chi}$, so that we do not derive
the condition for having a completely stable vacuum. Note however that 
even a local minimum is likely to be very long lived, barriers with 
sizes of order TeV implying lifetimes typically longer than the age of the 
universe \cite{Dasgupta:1997pz}.

\section{Light Boson Spectrum}
\setcounter{equation}{0}

In the previous section we have seen that the charged Higgs,
three of the CP-even neutral scalars and one CP-odd neutral scalar
are always heavy, with masses of order $\sqrt{\lambda} v/\epsilon$,
in the TeV range. The only remaining physical states are the 
CP-even $h^0$, and the CP-odd $A^0_{t}$ and $A^0_{\chi}$. Furthermore,
the vacuum stability condition implies that only one of
$A^0_{t}$ and $A^0_{\chi}$ may have a mass of order $v$ or smaller.
Therefore, there are three possible contents for the composite Higgs 
spectrum below a TeV scale: \\
1. Only the $h^0$;  \\
2.  the $h^0$ and $A^0_{t}$;  \\
3.  the $h^0$ and $A^0_{\chi}$. \\ 
In this Section we analyze these cases in turn.

\subsection{Standard Model in the decoupling limit}

If the only scalar lighter than a scale of order 1 TeV 
is the CP-even Higgs boson, $h^0$, then the low energy theory 
has precisely the Standard Model field content.
The corrections due to the heavier states are of order $\epsilon^2$.
Therefore, the Standard Model is obtained in the decoupling limit
where $\epsilon \ll 1$. 
However, in practice  $\epsilon$ cannot be smaller than one
by many orders of magnitude if we want to avoid an exponential fine-tuning.
Note that the mass terms in the effective potential have coefficients
of order $M_{H^\pm}^2$, which is  larger than $v^2$ by a factor $1/\epsilon^2$.
This means that the extremization conditions listed in the Appendix
require a fine-tuning of order $\epsilon^2$.

Due to the current agreement of the Standard Model 
to the experimental data, it follows that the Minimal Composite Higgs Model
discussed in this paper is viable for small $\epsilon$.
The strongest bound, $\epsilon \lae 0.2$ comes from the $\rho$ parameter, 
which receives corrections due to the $t-\chi$ mixing 
\cite{Dobrescu:1998nm,Chivukula:1998wd}. 
This bound is loose enough to avoid worrisome fine-tuning, but 
sufficient to make the decoupling limit a reasonable approximation.

Since the Standard Model is the decoupling limit of an underlying 
theory with dynamical electroweak symmetry breaking, the Higgs boson mass is 
a function of the parameters of the high energy theory.
Hence, one has to check whether there are restrictions on the Higgs boson mass
in addition to the usual Standard Model upper bounds from unitarity and 
triviality, and the lower bounds from direct searches. 
Note that the indirect upper bound on $M^2_h$
from the electroweak data is not constraining unless the scale of new physics 
is very high \cite{Chivukula:1999az}.
Also, the constraint from vacuum stability at large field is easily relaxed 
in the presence of new physics \cite{Datta:1996ni}.

From the expression for $M^2_{h^0}$ in eq.~(\ref{mhiggs}) it is clear that 
the upper end of the  Standard Model range can be reached when 
$x_t, \, x_\chi \gg 1$, which corresponds to large values for $\mu_{\chi t}$
and $\mu_{\chi\chi}$. 
By reducing $x_t$ and $x_\chi$ continuously we can cover the whole mass range 
of the Standard Model Higgs boson. 

It is useful to find out in more detail 
the situations in which $M_{h^0}$ may be as light as ${\cal O}(100)$ GeV.
To this end, we would like to express $M_{h^0}$ in terms of the parameters 
of the effective potential. 
For simplicity we will consider the ``seesaw limit'', $\tan\beta \gg 1$, in 
which only the $H_\chi$ doublet is responsible for the bulk of electroweak 
symmetry breaking, and the top mass is produced almost entirely
via the seesaw mechanism.
To leading order in $1/\tan\beta$ and $\epsilon^2$, 
the Higgs boson squared-mass takes the form
\be
M^2_{h^0} = \frac{2 v^2 }{M^2_{\chi\chi} - 3 M^2_{t\chi} }
  \left[ \xi^2 \left( M^2_{\chi\chi} - M^2_{t\chi} \right)
+ \frac{ 2 M^4_{t\chi}}{M^2_{\chi\chi} - 3 M^2_{t\chi}}
\left( 1 + \frac{ 4 M^4_{t\chi} }{M^2_{\chi t} -  M^2_{t\chi}} \right)
+ {\cal O}\left(\frac{1}{\xi^2} \right)  \right] ~.
\ee
In deriving this equation we have used $\cos\gamma \approx 1/\xi < 1$,
which follows from the expression (\ref{top-mass}) for the top quark mass,
and $\lambda = 2 \xi^2$. 
The leading order in $1/\xi^2$ has been derived previously in
\cite{Chivukula:1998wd}, 
where it is argued that there may be natural situations in which 
the underlying theory above the compositeness scale dictates
a partial cancellation between $M^2_{\chi\chi}$ and $M^2_{t\chi}$,
making a light composite Higgs boson a distinct possibility.
In practice it is sufficient that this cancellation is of order 
$1/\xi^2 \sim 10\%$. To see this, let us define a parameter 
$d_M \sim {\cal O}(1)$ by
\be
\frac{M^2_{\chi\chi}}{M^2_{t\chi}} =  1 + \frac{d_M}{\xi^2}~,
\ee
and assume for simplicity that $M^2_{\chi t} \gg M^2_{t \chi}$.
The $M^2_{h^0}$ dependence on $d_M$,
\be
M^2_{h^0} \approx \left(1- d_M \right) v^2  ~,
\ee
shows that the Higgs boson mass
can easily be below the electroweak scale in this case. 

One may wonder how large are the radiative corrections to the 
Higgs boson mass. In fact we have already included the leading large-$N_c$
loop corrections when we derived the effective potential. 
The corrections from the quartic and trilinear scalar couplings are 
in general significant given that the quartic coupling in the effective 
potential is large. However, these contributions are of order $1/N_c$
compared to the ones we included, and we will assume that their effects do
not change qualitatively our results. 

Although in the decoupling limit the low energy effective theory 
looks like the Standard Model, the Minimal Composite Higgs Model has 
a distinctive feature: the trilinear and quartic Higgs boson couplings
are large and rather independent of the Higgs boson mass. 
The quartic coupling is given by $\lambda/8$ while the trilinear coupling is 
$\sim \lambda v/2$. If the Higgs boson will be discovered, it is 
conceivable that its trilinear coupling will be measured at futures 
colliders \cite{Djouadi:1999rc}, 
and therefore the Minimal Composite Higgs Model
will be tested even if all other composite states happen to be 
heavier than the reach of those collider experiments.

\subsection{Light top-axion}

If the amount of $U(1)_t$ explicit symmetry breaking is small,
namely $C_{\chi t} \ll |\langle \phi_{\chi t} \rangle |^3$, 
then the $A^0_t$ is much lighter than the $H^\pm$. 
From Eq.~(\ref{param}) we find that the $A^0_t$ has a mass of the order of 
the electroweak scale or below for
\be
\mu_{\chi t} \lae \frac{v^3}{2 \pi \epsilon M_c^2} ~.
\ee
In the limit where $\mu_{\chi t} = 0$, the $A^0_t$   receives a small
mass only from the QCD anomaly. Although such an extreme case is ruled out
(see Section 6), we will refer to  $A^0_t$ as the ``composite top-axion'',
because it couples to the right-handed top.

The $h^0$ has a large trilinear coupling to $A^0_t$ pairs:
\be 
{\cal L}_3 \approx \frac{\lambda}{2} v a_t \;
h^0 \left(A^0_t\right)^2 ~.
\ee
At tree level, $a_t \sim \epsilon^2$ because to leading order in $\epsilon$,
the $h^0$ belongs to the  Higgs doublets, whereas the $A^0_t$ 
is part of the $\phi_{\chi t}$ singlet.
There are however large one-loop contributions, as shown in Fig.~3, 
due to the $\lambda \gg 1$ quartic coupling. These give $a_t \sim 1/N_c$.
The contributions from more loops which involve the quartic coupling
are suppressed by more powers of $1/N_c$, so are unlikely to change the 
order of magnitude of $a_t$.
\begin{figure}[t]
\centering
\begin{picture}(198,100)(-15,-6)
\put(101,50){\circle{42}}
\thicklines
\put(122,50){\line(4, -3){10}}\put(138,38){\line(4, -3){10}}
\put(154,26){\line(4, -3){10}}\put(170,14){\line(4, -3){10}}
\put(122,50){\line(4, 3){10}}\put(138,62){\line(4, 3){10}}
\put(154,74){\line(4, 3){10}}\put(170,86){\line(4, 3){10}}
\put(190,0){$A^0_t$}\put(190,86){$A^0_t$}
\put(80,50){\line(-2, 0){11}}
\put(63,50){\line(-2, 0){11}}
\put(46,50){\line(-2, 0){11}}
\put(29,50){\line(-2, 0){11}}
%
\put(20,35){$h^0$}
%
\end{picture}
\caption[]{
\label{Figure3}
\small One-loop contribution of the heavy scalars to the 
trilinear coupling of the Higgs boson to composite-axion 
pairs. }
\vspace{3mm}
\end{figure}
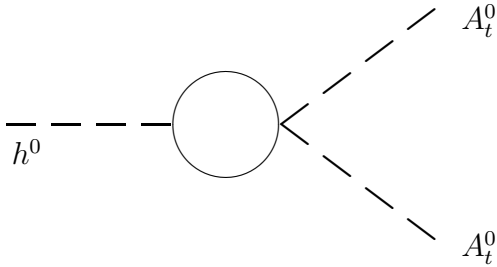

This large trilinear coupling is very important for Higgs boson searches
if $M_{h^0} > 2 M_{A^0_t}$.
The width for the Higgs boson decay into top-axion pairs,
\be
\Gamma \left(h^0 \rightarrow A^0_t A^0_t\right)
= \frac{\lambda^2 v^2 a_t^2}{32 \pi M_{h^0}}
\sqrt{1- \frac{4 M_{A^0_t}^2}{M_{h^0}^2}} ~,
\ee
is of the order of the Higgs mass for a light Higgs boson, and decreases
for larger $M_{h^0}$.

The Higgs boson mass has a simple form when the top-axion is light.
To show this, we remark that
$M_{A^0_t} \lae v \sim \epsilon M_{H^\pm}$
implies $x_t \lae \epsilon^2$. 
Then, using the expression for the top mass (\ref{top-mass})
with $y_t = m_t \sqrt{2}/v \approx 1$, we may write 
the Higgs boson squared-mass as
\be
M^2_{h^0} = \lambda v^2 \left[ 1 - \frac{1}{\cos^2\!\gamma} 
\left( \cos^2\!\beta + \frac{1}{x_\chi \xi^2}\right)
 + {\cal O}(1/\xi^2) \right] ~.
\ee
It appears that the full Standard Model range is open for the 
Higgs boson mass, but a light $h^0$ requires
$\cos\beta/\cos\gamma \sim 1$, or a fine-tuning of $x_\chi \approx 1$
({\it i.e.}, $M_{A^0_\chi} \approx 2 m_\chi$).
Therefore, the Higgs boson is generically a very broad resonance, which decays 
most of the time into top-axion pairs, or into $W$ and $Z$ pairs
for large $M_{h^0}$.

\subsection{Light $\chi$-axion}

The last possible light composite scalar is $A^0_\chi$.
This is similar with the light $A^0_t$ case: for 
small $\mu_{\chi \chi}$  the  amount of $U(1)_\chi$ explicit 
breaking is small and the $A^0_\chi$ may have a mass below 
the electroweak scale. We will call $A^0_\chi$ the 
``composite $\chi$-axion''.

Recall that there is no region of the parameter space in which 
both $A^0_t$ and $A^0_\chi$ are light. Therefore the only scalar
trilinear coupling relevant at current collider energies is
\be 
{\cal L}_3 = \frac{\lambda }{2} v a_\chi\, 
h^0 \left(A^0_\chi\right)^2 ~,
\ee
where the renormalized value of $a_\chi$ is again $\sim 1/N_c$.
If $M_{h^0} > 2 M_{A^0_\chi}$, then the branching ratio for 
the $h^0 \rightarrow A^0_\chi A^0_\chi$ decay mode may be large.
The width for the Higgs decay 
into a $\chi$-axion pair is similar with that from the light-$A^0_t$ case,
and can be estimated using the value of $\lambda$ from Eq.~(\ref{lambda}):
\be
\Gamma \left(h^0 \rightarrow A^0_\chi A^0_\chi\right)
\sim  \frac{8 \pi^4 v^2}{N_c^4 M_{h^0} \ln^2\left(M_c/m_\chi\right) }
\sqrt{1- \frac{4 M_{A^0_\chi}^2}{M_{h^0}^2}} ~.
\ee

Since the $h^0$ has Standard Model couplings to the weak gauge bosons
(because the other CP-even neutral scalars decouple up to $\epsilon^2$), 
we can immediately compare its widths
for the decays into $\chi$-axions and into $W$ or $Z$ pairs:
\be
\frac{\Gamma \left(h^0 \rightarrow A^0_\chi A^0_\chi\right)}
{\Gamma \left(h^0 \rightarrow WW, ZZ\right)} \approx 
\frac{ \lambda^2 a_\chi^2 v^4}{3 M_{h^0}^4}
~,
\ee
where we neglected the $\chi$-axion mass and the gauge boson masses.
The dominant decay mode  of a Higgs boson lighter than the electroweak scale
is into $\chi$-axion pairs.

The novel feature of the light $\chi$-axion case is that it places 
an upper bound on the Higgs boson mass.
The condition for a light $\chi$-axion,  
$M_{A^0_\chi} \lae \epsilon M_{H^\pm}$, implies 
$x_\chi \lae \epsilon^2$, and the Higgs boson squared-mass becomes
\be
M^2_{h^0} = v^2 \left[ 4 \xi \cos\beta - \frac{2}{x_t} 
+ {\cal O}\left(1/\xi^2\right) \right]  ~.
\ee
Because $\cos\beta < 1/\xi$, we find that the upper bound on
the Higgs boson mass is $2v$.
This bound is not very stringent, but still
relevant for searches at the LHC.

\section{Composite Scalar Couplings to Quarks and Leptons}

The couplings of the light bosons to the quarks and leptons 
are model dependent, as in a general two-doublet Higgs model.
All quarks and leptons have to couple to at least one of the two 
Higgs doublets in order to acquire masses. Such couplings may arise in 
the low energy effective theory in various ways, depending on the 
structure of the underlying theory above the compositeness scale.
A simple possibility is that there are four-fermion couplings between
the $t$, $\chi$ and the light fermions:
\be
\frac{1}{M_c^2} \left[
\left(\overline{\psi_L^j} u_R^k \right) 
\left( \eta^u_{jk}  \,  \overline{t_R} \psi_L^3 
+ \eta^{\prime u}_{jk}  \, \overline{\chi_R} \psi_L^3 \right) 
+ \left(\overline{\psi_L^j} d_R^l \right) i\sigma_2 
\left( \eta^d_{jl}  \, \overline{\psi_L^3 } t_R
+ \eta^{\prime d}_{jl}  \, \overline{\psi_L^3} \chi_R \right)\right] 
+ {\rm h.c.}
\ee
For brevity,
we show here only the four-fermion operators involving quarks. 
The couplings of $\overline{\chi}_R \psi_L^3$ and $\overline{t}_R \psi_L^3$
to the leptons have the same form.
The generational indices $j$ and $l$ run from 1 to 3, 
while $k$ runs from 1 to 4 because the $\chi_R$ may mix with 
the $u_R$, $c_R$ and $t_R$ weak eigenstates.
The above set of four-fermion
operators may be viewed as a parametrization of the flavor
symmetry breaking effects, whose origin could be explained in principle 
within a variety of high-energy theories, as discussed in section 2.3.
The four-fermion operators 
 give rise in the low energy theory to Yukawa couplings of the
$H_t$ and $H_\chi$ doublets to the quarks and leptons:
\be
- \left( \overline{\psi}_L^j u_R^k\right) \left( \lambda^u_{jk} H_t 
	+ \lambda^{\prime u}_{jk} H_\chi\right)
+ \left( \overline{\psi}_L^j d_R^l \right) i\sigma_2
\left( \lambda^d_{jl} H_t^\dagger 
	+ \lambda^{\prime d}_{jl} H_\chi^\dagger \right) ~.
\label{light-fermion}
\ee
\begin{figure}[t]
\centering
\begin{picture}(198,100)(-15,-6)
\thicklines
\put(100,50){\circle{40}}
\put(122,50){\line(4, -3){45}}
\put(122,50){\line(4, 3){45}}
\put(80,50){\line(-2, 0){11}}
\put(63,50){\line(-2, 0){11}}
\put(46,50){\line(-2, 0){11}}
\put(29,50){\line(-2, 0){11}}
\put(12,50){\line(-2, 0){11}}
\put(95,15){$\psi_L^3$}
\put(0,35){$H_t, \, (H_\chi)$}
\put(85,77){$t_R \; (\chi_R)$}
\end{picture}
\caption[]{
\label{Figure4}
\small  Leading-$N_c$ contribution to the 
Yukawa couplings of the Standard Model fermions.}
\vspace{3mm}
\end{figure}
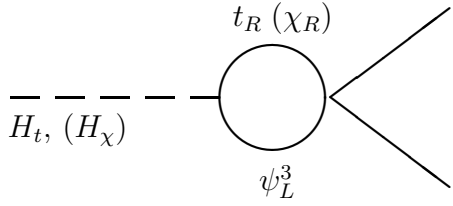
The Yukawa coupling constants, 
$\lambda^d, \lambda^{\prime d}, \lambda^u, \lambda^{\prime u}$,
are proportional with the coefficients of the four-quark operators,
$\eta^d, \eta^{\prime d}, \eta^u, \eta^{\prime u}$, respectively. 
The factor of proportionality may be estimated by computing the leading-$N_c$
contribution shown in Fig.~4, with a physical cut-off at $M_c$, and
the result is 
\be
\left(\lambda^d, \, \lambda^{\prime d}, \, \lambda^u, \, 
\lambda^{\prime u} \right) \approx
\frac{\xi N_c}{8 \pi^2}
\left(- \eta^d, \, \eta^{\prime d}, \, \eta^u, \, - \eta^{\prime u} \right) ~.
\ee

Note that six-fermion couplings and other higher-dimensional couplings
can also contribute to the light quark and lepton masses. 
Below the compositeness scale, they give rise to terms in the 
effective Lagrangian involving several Higgs doublets
and fermions. If they  give
the dominant contribution to some of the fermion masses, then 
the couplings of the Higgs boson to those fermions are non-standard
\cite{Babu:1999me}. 
Another possibility for fermion mass generation is to let
all quarks and leptons to participate in a seesaw mechanism,
by extending the vector-like quark sector 
\cite{family, Simmons:1989pu}.
In what follows we will ignore these possibilities, and study the 
couplings induced by the Yukawa couplings shown above.

Only the linear combination
$\, (H_t \cos\beta + H_\chi \sin\beta) \,$ has an electroweak asymmetric
VEV, so that the down-type quark masses are given by 
\be
{\rm diag}( m_d, m_s, m_b ) = \frac{v}{\sqrt{2}}  S_d^\dagger 
\left( \lambda^d \cos\beta
+ \lambda^{\prime d} \sin\beta \right)  T_d ~,
\label{down-mass}
\ee
where $S_d$ and $T_d$ are unitary matrices. A similar statement applies to 
the lepton sector.
The only light scalar contained in this linear combination is the $h^0$,
and its induced couplings to down-type quarks or leptons 
are Standard-Model-like up to corrections of order 
$\epsilon$.
Note that the $b$-quark mass requires $\eta^d_{33} \cot\beta + 
\eta^{\prime d}_{33} \sim 0.2$, which shows that generically the
coefficients of the four-quark operators responsible for light fermion 
masses are indeed perturbative at the compositeness scale.

The other  linear combination of Higgs doublets,
\be
- H_t \sin\beta +  H_\chi \cos\beta = 
\left( \begin{array}{c} \frac{1}{\sqrt{2}} \left(
+ h^0_{t\chi}\cos\beta - h^0_{tt}\sin\beta 
- i A^0 \right)
\\ [4mm] - H^- \end{array} \right) ~,
\label{no-VEV}
\ee
has couplings which may induce FCNC's in the down-type quark sector.
The charged Higgs induces FCNC's at one loop level, but is
sufficiently heavy to make these effects insignificant.
The neutral states, however contribute to FCNC's at tree level,
and we have to make sure that these contributions are not too large.
The couplings of the neutral scalars to the down-type quark mass eigenstates
are given by 
\bear
\frac{1}{\sqrt{2}} \overline{d}_L S_d^\dagger \left( - \lambda^d \sin\beta
+ \lambda^{\prime d} \cos\beta \right) T_d d_R
\left[ H^0_1 + i A^0 \right.
\nonumber \\ [3mm] \left.
+ \epsilon \left( - i c_A A^0_t + i c_A^\prime A^0_\chi 
+ c_H H^0_2 + c_H^\prime H^0_3\right) + {\cal O}(\epsilon^2) \right] ~,
\label{axion-coupling}
\eear
where $c_H$ and $c_H^\prime$ are parameters of order one in the $\epsilon$
expansion. Notice that the $h^0$ couplings are not affected by these terms
up to order $\epsilon^2$.

In general, these couplings may be flavor non-diagonal because the
FCNC's induced by them are suppressed at order $\epsilon^2$. 
However, in order 
to avoid too strong bounds on $\epsilon^2$ (which would correspond to
fine-tuning), it is preferable to assume that the matrix 
$S_d^\dagger \left( - \lambda^d \sin\beta
+ \lambda^{\prime d} \cos\beta \right) T_d$ is approximately flavor-diagonal. 
There are many situations in which this happens. For example, 
when the two matrices $\lambda^d$ and
$\lambda^{\prime d}$ are approximately proportional, or when one of them
vanishes.


The scalar couplings to up-type quarks are more complicated  
due to the mixing with the $\chi$.
The up-type quark mass matrix is $4 \times 4$, and has large elements 
corresponding to the $\chi$ and $t$ weak eigenstates, given by 
eq.~(\ref{f1}). The other elements are given by the Yukawa couplings 
(\ref{light-fermion}) and are typically small because they are produced
by perturbative four-quark operators at the $M_c$ scale or above.
Since the $\chi$ is much heavier than the electroweak scale, its mixing
with the quarks other than $t$ is small. If we ignore this mixing altogether,
we have a situation similar with that in the down-type sector:
the $h^0$ has Standard Model couplings up to corrections of order 
$\epsilon^2$, while the $A^0_t$ and $A^0_\chi$ have couplings of order
$\epsilon$ to the Standard Model fermions.
On the other hand, the mixing of the up-type quarks with the $\chi$ 
could lead to certain flavor non-diagonal couplings of the $h^0$ which may be
allowed by the FCNC constraints, while producing interesting phenomena
such as single-top decays of the Higgs boson \cite{Burdman:1999sr}.
Note also that the $h^0$ has a large coupling, of $\sim \xi/\sqrt{2}$,
to the $\overline{t}_L \chi_R$ quark mass eigenstates.

\section{Bounds on the Composite Axion}
\setcounter{equation}{0}

In this section we study the lower mass bounds on the composite axion.
These are sensitive to the axion couplings to fermions, 
which are of order $\epsilon$ or smaller, and depend on its identity
($A_t^0$ or $A_\chi^0$) only up to an overall 
constant, as can be seen from eq.~(\ref{axion-coupling}).
The axion-fermion couplings are very model dependent
and is beyond the scope of this paper to comprehensively analyze 
the mass bounds in all cases. We will rather 
concentrate on the cases which are most favorable for a light axion.

The tree level axion couplings to light quarks and leptons have two sources.
One of them is the Yukawa couplings to light fermions of the 
 doublet with no VEV, shown in Eq.~(\ref{no-VEV}).
These may vanish or be very small because they are
 not restricted by the quark and lepton masses.
The other source is the Yukawa couplings of the composite scalars to their 
constituents, Eq.~(\ref{yukc}). After transforming to the mass eigenstate
basis,
these Yukawa interactions induce  axion couplings to all the up-type quarks.
However, the mixings between the weak eigenstates of the 
$t$ or $\chi$ with $u$ and $c$ are again unrestricted, and may vanish
without affecting the CKM matrix. Notice that in this case the
 $V_{ts}$ and $V_{td}$ elements
are fixed by the $S_d$ unitary matrix [see Eq.~(\ref{down-mass})],
while $V_{ub}$ and $V_{cb}$ are combinations of the transformations
in both the up- and down-type sectors.
Once a predictive and compelling theory of flavor is found, one
can decide whether the aforementioned mixings and couplings are naturally
small or vanishing. 

Here we will assume they do, so that  the only fermions that 
couple at tree level to the composite 
axion ($A_t^0$ or $A_\chi^0$) are
the $t$ and $\chi$ mass eigenstates:
\begin{equation}
\frac{i \xi}{\sqrt{2}}
\left(\overline{t_L} ,\,\overline{\chi_L}\right)\,
\left[ A^0_t
\pmatrix{ {\cal O}(\epsilon)  & {\cal O}(\epsilon)  \cr 
1 & 0 \cr} 
+ A^0_\chi 
\pmatrix{ {\cal O}(\epsilon)  & {\cal O}(\epsilon) \cr 0 & 1 \cr} 
+ {\cal O}(\epsilon^2)
\right] \Gamma
\pmatrix{- t_R \cr \chi_R \cr} +{\rm h.c.} ~,
\label{axion}
\end{equation}
where $\Gamma$ is the unitary matrix given in Eq.~(\ref{Gamma}).
Therefore, the composite axion may be produced at colliders through 
a $t$ or $\chi$ loop, but the production rate is too small for placing bounds
even at the $Z$ pole at LEP \cite{Randall:1992gp}. 

The quarkonium decays could in principle constrain the composite axion mass.
However, the current limit on the branching ratio of the
most promising decay mode, 
$\Upsilon(1S) \rightarrow A^0_{t, \, \chi} \gamma$, is at the level of 
$10^{-5}$ \cite{Caso:1998tx}, which is not sufficient for constraining the
composite axion. Note that this decay occurs through a top-loop, and 
a  suppression factor of order $\epsilon^2$ appears in the width.

The $K^+  \rightarrow A^0_{t, \, \chi} \pi^+$ decay is another usual
suspect for constraining the axions. This again involves a top-loop 
and is further suppressed by $V_{ts}V_{td}$, so that no useful 
mass bounds can be derived.

More generally, if the axion is coupled at tree level only to the $\chi$
and $t$, it is sufficiently insulated from the light fermions to 
avoid constraints from usual laboratory searches. 
The astrophysical constraints are harder to avoid. The composite axion
may be produced in stars if it is light enough, leading to unacceptable 
cooling rates. At one-loop, the axion couples to gluon pairs and to 
photon pairs. Combined, these couplings rule out very light axions with a decay
constant below $\sim 10^{10}$ GeV
\cite{Caso:1998tx}. In our model the axion decay 
constant is of order $v/\epsilon$, implying that a very small value for 
$\epsilon$ is required, which leads to an exponential fine-tuning. 
Note that in the limit where the $\mu_{\chi t}$ or $\mu_{\chi \chi}$
mass parameter vanishes, 
the only contribution to the 
axion mass is given by the QCD anomaly, and 
it is tempting to solve the strong CP problem using the composite axion.
However, the small value for $\epsilon$ is not encouraging.
We therefore do not attempt to solve the strong CP problem, and assume 
that the Peccei-Quinn symmetry is explicitly broken by a non-zero 
$\mu_{\chi t}$ and $\mu_{\chi \chi}$, or by some higher dimensional 
operators.

The bound on the axion decay constant is avoided if the  composite axion is 
heavier than the 
core temperature of the stars by an order of magnitude, 
because the axion production is Boltzmann suppressed
and the cooling rate is not much affected.
The red giant stars have a core temperature of order 10 keV which impose a
lower mass bound of $\sim 200$ keV on the axion \cite{Kolb:1990vq}.

The larger temperature of the supernova 1987A, about 30 MeV at the center,
appears to yield the most 
stringent lower limit on the composite axion mass.
However, the  cooling rate of the  supernova is not affected by our 
composite axion, because the very 
high density of the newly formed neutron star reduces the 
axion emission to acceptable levels for an
 axion decay constant below $\sim 10^6$ GeV \cite{Kolb:1990vq}.
Although the axion flux could not affect the supernova cooling,
there are constraints due to the absence of an axion signal in
water Cerenkov detectors during the SN 1987A \cite{Engel:1990zd}. These impose 
a lower bound of a few hundred TeV on the axion decay constant.
We find more reasonable to evade this bound by imposing a lower
limit on the composite axion mass of ${\cal O}(100)$ MeV, such that 
its production is substantially suppressed.

If $A_t^0$ or $A_\chi^0$ has a mass between this lower bound and 
2$m_\pi \approx 270$ MeV, then the $\pi^0\pi^0$ decay channel is closed, 
and the 
composite axion decays predominantly to photon pairs (assuming that 
the tree level coupling to $e^+ e^-$ vanishes).
In this case, the CP-even Higgs boson, which decays  most of the
time to axion pairs, will have a striking signature at future 
colliders: two pairs of almost collinear photons. 

\section{Discussion}

It is instructive to make a comparison of the Minimal Composite Higgs Model
(MCHM) presented here with the Minimal Supersymmetric Standard Model (MSSM).
Both these models have a decoupling limit in which they look like the 
Standard Model, and therefore are consistent with current electroweak 
precision data. Both models include two Higgs doublets, but the 
composite model requires also two gauge singlet fields resulting in a 
more complicated Higgs sector. The top quark plays an active role 
in electroweak symmetry breaking within both the MCHM and MSSM.

These two models may be viewed as effective theories whose parameters 
have to be determined by higher-energy physics. The MCHM includes four 
coefficients of the four-quark operators which are fixed by the 
gauge couplings and representations of the top and $\chi$ quarks, and possibly 
by their position in extra dimensions. The MSSM has soft supersymmetry breaking
parameters which need to be determined within a theory of dynamical 
supersymmetry breaking. Similarly, the presence of the gauge invariant 
fermion mass terms that are present in the MCHM, and the $\mu$ term in the 
MSSM are hopefully accounted for by physics at higher energy.

From a more theoretical point of view, both the MCHM and the MSSM
can be linked to low energy manifestations of 
certain features that are likely to occur in a more comprehensive theory 
which includes quantum gravity, such as string or M theory. 
Furthermore, in the presence of extra dimensions compactified at a 
scale in the TeV range, the composite Higgs model is compatible with
gauge coupling unification \cite{Cheng:1999fu}, 
although this cannot be checked at the level
of precision allowed by the perturbativity of the MSSM.

Despite these similar aspects, the MCHM and MSSM are conceptually 
different. In the MCHM there is no fundamental Higgs field.
Therefore, the origin of electroweak symmetry breaking is found
in dynamical phenomena, as opposed to the radiative corrections
involved in the MSSM.
Also, the phenomenology of the MCHM and MSSM is different.
In the MCHM there are no superpartners at the electroweak scale, 
but there is a potentially light axion, a heavy vector-like quark,
and interesting phenomena at scales in the TeV range,
associated with the strong dynamics.

An important phenomenological aspect of the MCHM is the dominant branching 
ratio (if allowed kinematically) of the Higgs boson decay into composite 
axion pairs. The discovery of the Higgs boson in this decay mode would 
be a spectacular evidence for the MCHM. On the other hand, if only a light
Standard Model Higgs boson will be discovered, it will probably be 
necessary to measure its trilinear coupling or to 
experiment with colliders at higher energies in order to 
distinguish between the MCHM, the MSSM, or other models with a decoupling 
limit. 

\vspace{3mm}

{\it Acknowledgements:} I would like to thank Bill Bardeen, Gustavo Burdman,
Hsin-Chia Cheng, Scott Dodelson,
Hong-Jian He,  Chris Hill, and Konstantin Matchev for useful discussions.

{\it Postscript:} 
A related study of a Top Quark Seesaw Model has appeared while 
this work was concluded \cite{new}. The focus of that study is rather
different than in this paper. For example, the models discussed there
do not have a decoupling limit in which the Standard Model with a light
Higgs boson is recovered.

\section*{Appendix: \ Extremization Conditions for the Effective Potential}
\addcontentsline{toc}{section}{Appendix:  \ Extremization Conditions for the 
Effective Potential} \renewcommand{\theequation}{A.\arabic{equation}}
\setcounter{equation}{0}

In this Appendix we list the extremization conditions for
the effective potential studied in Section 3. These can be read
from Eq.~(\ref{efpot}) by imposing the cancellation of the 
tadpole terms:
\bear
\hspace {-5mm} \frac{\partial V(0)}{\partial h^0_{tt}} 
\hspace{-2mm} & = & \hspace{-2mm} 
	v\cos\beta \left[ M_{tt}^2 + 
\frac{\lambda v^2}{2 \epsilon^2} 
\left( r_s \frac{\tan\beta}{\tan\gamma} + \epsilon^2 \right)\right] = 0
\nonumber \\ [3mm]
\hspace{-5mm} \frac{\partial V(0)}{\partial h^0_{t\chi}}
\hspace{-2mm}  & = &  \hspace{-2mm} 
	v\sin\beta \left[ M_{t\chi}^2 - \frac{\lambda v^2}{2 \epsilon^2} 
\left( r_s  - \epsilon^2 \right)\right] = 0
\nonumber \\ [3mm]
\hspace{-5mm} \frac{\partial V(0)}{\partial h^0_{\chi t}} 
\hspace{-2mm} & = & \hspace{-2mm} 
	- \frac{v \sin\beta}{\epsilon \tan\gamma} \left[ M_{\chi t}^2 + 
\frac{\lambda v^2}{2 \epsilon^2} 
\left( \frac{\sin^2\!\beta}{\sin^2\!\gamma} 
+ \epsilon^2 r_s \frac{\tan\gamma}{\tan\beta}  
 \right) \right] + \sqrt{2} \, C_{\chi t} = 0 
\nonumber \\ [3mm]
\hspace*{-5mm} \frac{\partial V(0)}{\partial h^0_{\chi\chi}}
\hspace{-2mm}  & = & \hspace{-2mm} 
	- \frac{v}{\epsilon} \sin\beta \left[ M_{\chi\chi}^2 + 
\frac{\lambda v^2}{2 \epsilon^2} 
\left(\frac{\sin^2\!\beta}{\sin^2\!\gamma}  - \epsilon^2 r_s 
\right) \right] + \sqrt{2} \, C_{\chi\chi} = 0 
\eear
where we introduced for convenience the following notation:
\be
r_s \equiv \frac{\sin\beta}{\sin\gamma} \cos (\beta + \gamma)~.
\ee
Note that $M_{t \chi}^2 < 0$ requires $\cos (\beta + \gamma) < 0$, while
the expression for the top mass (\ref{top-mass}) imposes
\be
-1 < r_s < -1 + \frac{1}{\xi^2} + {\cal O} (1/\xi^4) ~.
\ee


\vfil

\begin{thebibliography}{99}
\frenchspacing



\bibitem{Kaplan:1984sm}
D.B.~Kaplan and H.~Georgi,
``SU(2) X U(1) Breaking By Vacuum Misalignment,''
Phys. Lett. {\bf 136B}, 183 (1984); \\
D.B.~Kaplan, H.~Georgi and S.~Dimopoulos,
``Composite Higgs Scalars,''
Phys. Lett. {\bf 136B}, 187 (1984); \\
T.~Banks,
``Constraints On SU(2) X U(1) Breaking By Vacuum Misalignment,''
Nucl. Phys. {\bf B243}, 125 (1984); \\
H.~Georgi, D.B.~Kaplan and P.~Galison,
``Calculation Of The Composite Higgs Mass,''
Phys. Lett. {\bf 143B}, 152 (1984); \\
H.~Georgi and D.B.~Kaplan,
``Composite Higgs And Custodial SU(2),''
Phys. Lett. {\bf 145B}, 216 (1984); \\
M.J.~Dugan, H.~Georgi and D.B.~Kaplan,
``Anatomy Of A Composite Higgs Model,''
Nucl. Phys. {\bf B254}, 299 (1985); \\
V.~Koulovassilopoulos and R.S.~Chivukula,
``The Phenomenology of a nonstandard Higgs boson in W(L) W(L) scattering,''
Phys. Rev. {\bf D50}, 3218 (1994)
hep-ph/9312317.

\bibitem{Bardeen:1990ds}
W.A.~Bardeen, C.T.~Hill and M.~Lindner,
``Minimal Dynamical Symmetry Breaking Of The Standard Model,''
Phys. Rev. {\bf D41}, 1647 (1990).

\bibitem{Nambu} Y.~Nambu,
``BCS Mechanism, Quasi Supersymmetry, And Fermion Masses'',
 in the Proceedings of the 
{\em XI Warsaw Symposium on Elementary Particle Physics, May 1988},
ed. Z.~Ajduk, {\it et al} (World Scientific, 1989); \
``Quasisupersymmetry, Bootstrap Symmetry Breaking And Fermion Masses,''
in the Proceedings of the {\em 1988 International Workshop on New Trends in 
Strong Coupling Gauge Theories}, Nagoya, Japan, ed. Bando, Muta and Yamawaki
(World Scientific, 1989); \
``Bootstrap Symmetry Breaking In Electroweak Unification'',
EFI-89-08 (1989); \\
V.A.~Miransky, M.~Tanabashi and K.~Yamawaki,
``Is The T Quark Responsible For The Mass Of W And Z Bosons?,''
Mod. Phys. Lett. {\bf A4}, 1043 (1989); \ 
``Dynamical Electroweak Symmetry Breaking With Large Anomalous Dimension And
                  T Quark Condensate,''
Phys. Lett. {\bf B221}, 177 (1989); \\ 
W.J.~Marciano,
``Heavy Top Quark Mass Predictions,''
Phys. Rev. Lett. {\bf 62}, 2793 (1989).

\bibitem{Cvetic:1997eb}
For a review, see G.~Cvetic,
``Top quark condensation: A Review,''
Rev. Mod. Phys. {\bf 71}, 513 (1999)
hep-ph/9702381.

\bibitem{Dobrescu:1998nm}
B.A.~Dobrescu and C.T.~Hill,
``Electroweak symmetry breaking via top condensation seesaw,''
Phys. Rev. Lett. {\bf 81}, 2634 (1998)
hep-ph/9712319.

\bibitem{Chivukula:1998wd}
R.S.~Chivukula, B.A.~Dobrescu, H.~Georgi and C.T.~Hill,
``Top quark seesaw theory of electroweak symmetry breaking,''
Phys. Rev. {\bf D59}, 075003 (1999)
hep-ph/9809470.

\bibitem{family}
G.~Burdman and N.~Evans,
``Flavor universal dynamical electroweak symmetry breaking,''
Phys. Rev. {\bf D59}, 115005 (1999)
hep-ph/9811357.

\bibitem{Dobrescu:1998dg}
B.A.~Dobrescu,
``Electroweak symmetry breaking as a consequence of compact dimensions,''
hep-ph/9812349.

\bibitem{Dobrescu:1999cs}
B.A.~Dobrescu,
``Higgs compositeness from top dynamics and extra dimensions,''
hep-ph/9903407.


\bibitem{Bando:1991rr}
M.~Bando, T.~Kugo, N.~Maekawa, N.~Sasakura, Y.~Watabiki and K.~Suehiro,
``Compositeness Condition In Renormalization Group Equation,''
Phys. Lett. {\bf B246}, 466 (1990).


\bibitem{topcolor}
C.T.~Hill,
``Topcolor: Top quark condensation in a gauge extension of the standard
                  model,''
Phys. Lett. {\bf B266}, 419 (1991).

\bibitem{Chivukula:1995dc}
R.S.~Chivukula, B.A.~Dobrescu and J.~Terning,
``Isospin breaking and fine tuning in topcolor assisted technicolor,''
Phys. Lett. {\bf B353}, 289 (1995)
hep-ph/9503203.

\bibitem{Chivukula:1998uf}
R.S.~Chivukula and H.~Georgi,
``Large N and vacuum alignment in topcolor models,''
Phys. Rev. {\bf D58}, 075004 (1998)
hep-ph/9805478.

\bibitem{Burdman:1999us}
G.~Burdman, R.S.~Chivukula and N.~Evans,
``Precision bounds on flavor gauge bosons,''
hep-ph/9906292.

\bibitem{Collins:1999cf}
H.~Collins, A.~Grant and H.~Georgi,
``Dynamically broken topcolor at large N,''
hep-ph/9907477.

\bibitem{largedim}
N.~Arkani-Hamed, S.~Dimopoulos and G.~Dvali,
``The Hierarchy problem and new dimensions at a millimeter,''
Phys. Lett. {\bf B429}, 263 (1998)
hep-ph/9803315; \\
I.~Antoniadis, N.~Arkani-Hamed, S.~Dimopoulos and G.~Dvali,
``New dimensions at a millimeter to a Fermi and superstrings at a TeV,''
Phys. Lett. {\bf B436}, 257 (1998)
hep-ph/9804398.

\bibitem{Randall:1999ee}
L.~Randall and R.~Sundrum,
``A Large mass hierarchy from a small extra dimension,''
hep-ph/9905221.
\bibitem{Lykken:1999nb}
J.~Lykken and L.~Randall,
``The Shape of gravity,''
hep-th/9908076.

\bibitem{Cheng:1999fu}
H.~C.~Cheng, B.A.~Dobrescu and C.T.~Hill,
``Gauge coupling unification with extra dimensions and gravitational scale
                  effects,''
hep-ph/9906327.

\bibitem{Arkani-Hamed:1999dc}
N.~Arkani-Hamed and M.~Schmaltz,
``Hierarchies without symmetries from extra dimensions,''
hep-ph/9903417.

\bibitem{Berezhiani:1998wt}
N.~Arkani-Hamed and S.~Dimopoulos,
``New origin for approximate symmetries from distant breaking in extra
                  dimensions,''
hep-ph/9811353; \\
Z.~Berezhiani and G.~Dvali,
``Flavor violation in theories with TeV scale quantum gravity,''
Phys. Lett. {\bf B450}, 24 (1999)
hep-ph/9811378.

\bibitem{extra}
K.R.~Dienes, E.~Dudas and T.~Gherghetta,
``Grand unification at intermediate mass scales through extra dimensions,''
Nucl. Phys. {\bf B537}, 47 (1999)
hep-ph/9806292; \\
S.A.~Abel and S.F.~King,
``On fixed points and fermion mass structure from large extra dimensions,''
Phys.\ Rev.\ {\bf D59}, 095010 (1999)
hep-ph/9809467; \\
A.B.~Kobakhidze,
``The Top quark mass in the minimal top condensation model with extra
                  dimensions,''
hep-ph/9904203; \\
K.~Yoshioka,
``On fermion mass hierarchy with extra dimensions,''
hep-ph/9904433.

\bibitem{Bando:1999di}
M.~Bando, T.~Kugo, T.~Noguchi and K.~Yoshioka,
``Brane fluctuation and suppression of Kaluza-Klein mode couplings,''
hep-ph/9906549.


\bibitem{Dasgupta:1997pz}
I.~Dasgupta, B.A.~Dobrescu and L.~Randall,
``Vacuum instability in low-energy supersymmetry breaking models,''
Nucl. Phys. {\bf B483}, 95 (1997)
hep-ph/9607487.


\bibitem{Chivukula:1999az} 
R.S.~Chivukula and N.~Evans,
``Triviality and the precision bound on the Higgs mass,''
hep-ph/9907414; \\
L.~Hall and C.~Kolda,
``Electroweak symmetry breaking and large extra dimensions,''
Phys. Lett. {\bf B459}, 213 (1999)
hep-ph/9904236; \\
R.~Barbieri and A.~Strumia,
``What is the limit on the Higgs mass?,''
hep-ph/9905281; \\
J.A.~Bagger, A.F.~Falk and M.~Swartz,
``Precision observables and electroweak theories,''
hep-ph/9908327.

\bibitem{Datta:1996ni}
A.~Datta, B.L.~Young and X.~Zhang,
``Implications of a nonstandard light Higgs boson,''
Phys. Lett. {\bf B385}, 225 (1996)
hep-ph/9604312.

\bibitem{Djouadi:1999rc}
A.~Djouadi, W.~Kilian, M.~Muhlleitner and P.M.~Zerwas,
``Testing Higgs selfcouplings at e+ e- linear colliders,''
Eur. Phys. J. {\bf C10}, 27 (1999)
hep-ph/9903229; \
``Production of neutral Higgs boson pairs at LHC,''
Eur. Phys. J. {\bf C10}, 45 (1999)
hep-ph/9904287; \\
D.J.~Miller and S.~Moretti,
``Can the trilinear Higgs selfcoupling be measured at future linear
                  colliders?,'' hep-ph/9906395.


\bibitem{Babu:1999me}
K.S.~Babu and S.~Nandi,
``Natural fermion mass hierarchy and new signals for the Higgs boson,''
hep-ph/9907213.

\bibitem{Simmons:1989pu}
E.H.~Simmons,
``Separating Electroweak Symmetry Breaking From Flavor Physics In An Almost
                  Standard Model,''
Nucl. Phys. {\bf B324}, 315 (1989).

\bibitem{Burdman:1999sr}
G.~Burdman,
``Scalars from top condensation models at hadron colliders,''
hep-ph/9905347.


\bibitem{Randall:1992gp}
L.~Randall and E.H.~Simmons,
``Signatures of neutral pseudoGoldstone bosons from technicolor,''
Nucl. Phys. {\bf B380}, 3 (1992); \\
G.~Rupak and E.H.~Simmons,
``Limits on pseudoscalar bosons from rare Z decays at LEP,''
Phys. Lett. {\bf B362}, 155 (1995)
hep-ph/9507438.

\bibitem{Caso:1998tx}
C.~Caso {\it et al.},
``Review of particle physics. Particle Data Group,''
Eur. Phys. J. {\bf C3}, 1 (1998).

\bibitem{Kolb:1990vq}
E.W.~Kolb and M.S.~Turner,
``The Early Universe,''
{\it  Redwood City, USA: Addison-Wesley (1990) 547 p. (Frontiers in physics,
                  69)}.

\bibitem{Engel:1990zd}
J.~Engel, D.~Seckel and A.C.~Hayes,
``Emission And Detectability Of Hadronic Axions From Sn1987a,''
Phys. Rev. Lett. {\bf 65}, 960 (1990).

\bibitem{new} 
 H.~Collins, A.~Grant, and H.~Georgi, 
 ``The Phenomenology of a Top Quark Seesaw Model,''
hep-ph/9908330.    

\end{thebibliography}
\end{document}